\documentclass[a4paper,12pt]{article}
\usepackage{graphicx}
\usepackage{amssymb}
\usepackage{amsmath}
\usepackage{cite}
\usepackage{multicol}
\def\be{\begin{eqnarray}}
\def\ee{\end{eqnarray}}

\title{Regularization of $1/X^2$ potential in general case of deformed space with minimal length }
\author{M. I. Samar and V. M. Tkachuk\\ Department for Theoretical Physics, \\ Ivan Franko National University of Lviv,\\ 12 Drahomanov St, Lviv,
UA-79005, Ukraine}
\begin{document}

\maketitle

\begin{abstract}
In general case of deformed Heisenberg algebra leading to the minimal length we present a definition of  the square inverse position  operator. 
Our proposal is based on the functional analysis of the square position operator.  Using this definition a particle in the field of the square inverse position  potential  is studied. We have  obtained analytical and numerical solutions for  the energy spectrum of the considerable problem in different cases of deformation function. We find that the energy spectrum slightly depends on the choice of deformation function.



\end{abstract}

\section{Introduction}
Idea of the minimal length have been attracted a lot of attention recently.
 This interest was inspired by the studies in  string theory and quantum gravity,
which suggest the existence of the minimal length as a finite lower bound 
to the possible resolution of length \cite{GrossMende,Maggiore,Witten}.
Minimal length can be achieved by modifying usual canonical
commutation relations \cite{Kempf1994,KempfManganoMann,HinrichsenKempf,Kempf1997}.
The most popular  deformed algebra is the one proposed by Kempf  \cite{Kempf1994}
\be \label{Kempf}
 [\hat{X},\hat{P}]=i\hbar (1+\beta \hat{P}^2),
\ee
leading to minimal length $\hbar\sqrt{\beta}$.

The one of the important problem concerning to minimal length hypothesis is to find quantum-mechanical (or classical) system with high sensitivity to deformation with minimal length.  The purpose is  not only in the estimation of the upper bound of minimal length but is in proving or refuting of the hypothesis of minimal length on the basis of comparison with experimental data.

In deformed space with minimum length coordinate representation does not exist.  Due to this fact, it is especially interesting to study the effect of the minimum length on systems with singular potentials, since such systems are expected to have the  nontrivial sensitivity to minimal length.

 The implications of non-zero minimal length were considered in the context of the following   problems with singularity in potential energy:   hydrogen atom \cite{Brau,Benczik,StetskoTkachuk,Stetsko2006,Stetsko2008, SamarTkachuk, Samar}, gravitational quantum well \cite{Brau2006, Nozari2010, Pedram2011}, a particle in delta potential and double delta potential\cite{Samar1, Ferkous}, one-dimensional Coulomb-like problem \cite{Samar1,Fityo,Samar2}, particle in the singular inverse square potential \cite{Bouaziz2007,Bouaziz2008, Bouaziz2017}.

Particle in the field  inverse square potential $1/R^2$ considered in paper \cite{Bouaziz2007,Bouaziz2008, Bouaziz2017} seems to be espesially interesting. The main feature of this system is the fact that such a system has no bound states in the undeformed case but do have the ones in case of deformed algebra with minimal length. In present paper we consider particle in one dimensional inverse square potential $1/X^2$ in general case of deformed algebra. The main aim this work is to study the dependence of the spectrum of the system on the choice of deformed algebra.

The paper is organized as follows. In section II we brief about general deformed algebra. In section III  we present  the definition of the $1/X^2$ operator based on the functional analysis of square position operator.
 In section IV    Schr\"odinger equation in momentum representation for the particle in $1/X^2$ potential is reduced to the solving of the differential equation  with boundary conditions. Section V contains analysys of energy spectrum for considered system in some particular cases of deformation. 
 Finally, in section V we conclude obtained results.

\section{ Deformed algebras and minimal length}
Let us  consider  a  modified  one-dimensional  Heisenberg  algebra which is generated  by
position $\hat{X}$ and momentum $\hat{P}$ hermitian operators satisfying the following relation
\be \label{general_deformation}
 [\hat{X},\hat{P}]=i\hbar f({\hat{P}}),
\ee
where $f$ are called functions of deformation. We assume that it is strictly positive ($f >0$), even function.

We consider a representation leaving position operator undeformed
\be\label{psevdo-position}
&&{\hat{X}}=\hat{x}=i\hbar\frac{d}{dp},\\ \nonumber
&&{\hat{P}=g({p})}.
\ee
The position and momentum operators in representation (\ref{psevdo-position}) act on square integrable function $\varphi(p) \in \it{L}^2(-b,b), (b\leq \infty)$. The
norm of $\varphi$ is given by
\be
||\varphi||^2=\int_{-b}^{b}{dp|\varphi(p)|^2}.
\ee

From the fact that operators $\hat{X}$ and $\hat{P}$, written in representation (\ref{psevdo-position}), have to satisfy the commutation relation (\ref{general_deformation}), we obtain the following differential equation for  $g(p)$ 
\be\frac{dg(p)}{dp}=f(P), \ee
with $P=g(p).$
Function $g({p})$  is an odd function  defined on $[-b,b]$, with $b=g^{-1}(a)$. Here $a$ represents the limits of momentum $P \in [-a,a]$.
 Minimal length for the deformed algebra is \cite{Maslowski}
\be l_0=\frac{\pi\hbar}{2b}.\label{minimal_length1} \ee
Thus, if $b<\infty$ nonzero minimal length exists and if $b=\infty$ the minimal length is zero. 
Here it is important to note that the result  obtained in \cite{Maslowski} concerning the minimal length depends on the choice of boundary condition for wavefunction. 
 Formula (\ref{minimal_length1}) was derived with assumption of  zero boundary conditions 
\be
\varphi(-b)=\varphi(b)=0.
\ee
Alternative choice of the boundary conditions may lead to elimination of the minimal length even for finite $b$ \cite{Nowicki}.

\section{Definition of $1/X^2$ operator}
To define square inverse distance operator  let us consider the eigenproblem for  $\hat{X}^2$
\be \label{X^2}
\hat{X}^2\phi_n(p)=\chi_n\phi_n(p).
\ee 
In representation (\ref{psevdo-position})  the square  distance operator has the form  $\hat X^2=-\hbar^2\frac{d^2}{dp^2}$ and acts on the domain
\be \label{DomainX2}
D(\hat X^2)=\left\{\varphi(p), \varphi''(p) \in{ \it{L^2}} \left(-b, b\right) , \ \varphi\left(-b\right)=\varphi(b)=0\right\}.
\ee
Now the eigenproblem (\ref{X^2}) writes   
\be
-\hbar^2\frac{d^2\phi_n(p)}{dp^2}=\chi_n\phi_n(p),
\ee
with $\phi_n(\pm b)=0$.

Note, that this problem is rather similar to the problem of particle in a box in undeformed space.
The solution of the problem is 
\be &&\chi_n=\frac{\hbar^2\pi^2}{4b^2} n^2, \label{ev}\\
&&\phi_n(p)=\frac{1}{\sqrt{b}}\sin\left(n\frac{\pi}{2b}(p+b)\right), \label{ef} \ee
with $n=1,2,\dots \ .$

We define the square inverse distance operator as 
\be
\frac{1}{\hat{X}^2}=\sum_{n=1}^{\infty}|\phi_n\rangle\frac{1}{\chi_n}\langle \phi_n|.
\ee
From this definition we see that the action of  the square inverse distance operator  ${1}/{\hat{X}^2}$ on any function belonging to its domain can be presented by
\be\label{X^-2}
\frac{1}{\hat{X}^2}\phi(p)=\int_{-b}^{b}dp'\phi(p')K(p,p').
\ee
with 
\be\label{kernel}
K(p,p')=\sum_{n=1}^{\infty}\phi_{n}(p)\frac{1}{\chi_{n}}\phi^*_{n}(p').
\ee
being the kernel of square inverse distance operator .
Substituting the expression for $\chi_n$  and $\phi_n(p)$  from  (\ref{ev}) and (\ref{ef}) into (\ref{kernel}) and using product-to-sum identity for sine functions we write
\be
K(p,p')=\frac{2b}{h^2\pi^2}\sum_{n=1}^{\infty}\frac{1}{n^2}\left(\cos\left[\frac{\pi n}{2b}(p-p')\right] -\cos\left[\frac{\pi n}{2b}(p+p'+2b)\right] \right).
\ee
Using the following formula\cite{Abramovic}
\be\sum_{n=1}^{+\infty}\frac{cos(n\theta)}{n^2}=\frac{\pi}{6}-\frac{\pi\theta}{2}+\frac{\theta^2}{4}, \  \ \theta\in[0,2\pi] \ee
we can finally write the kernel of square inverse distance operator in the following form
\be\label{kernel_1}
K(p,p')=-\frac{1}{2\hbar^2}\left(|p-p'|+\frac{pp'}{b}-b\right).
\ee
Thus, (\ref{X^-2}) together with (\ref{kernel_1})  presents  the definition of the square inverse distance operator in representation (\ref{psevdo-position}).
Note, that $K(\pm b,p')=0$, which means that the action of the square inverse distance operator
returns the wavefunction belonging to the domain of square distance operator (\ref{DomainX2}).

\section {$1/\hat X^2$ quantum well and minimal length}
The  Schr\"odinger equation for the particle in potential $-\alpha/\hat{X}^2$ with minimal length can be written as
\be \label{Shroedinger}
\left(\frac{1}{2m}g^2(p)-E\right)\psi(p)-\alpha\int_{-b}^{b}K(p,p')\psi(p')dp'=0, 
\ee
where $K(p,p')$ is given by (\ref{kernel_1}).
By changing $p$ for $-p$ in (\ref{Shroedinger}) we can show that the Hamiltonian commutes with
the parity operator, thus its eigenfunctions can be chosen as even or odd functions,
i.e. as parity eigenstates.

Differentiating twice the Shr\"odinger equation we obtain
\be \label{Shroedinger_diff} F''(p)=-\frac{2m\alpha}{\hbar^2}\psi(p),\ee
with $F(p)=\left(g^2(p)-2mE\right)\psi(p).$

Expressing $\psi(p)$ from (\ref{Shroedinger_diff}) and substituting it into (\ref{Shroedinger}) we obtain that both even and odd  eigenfunctions have to satisfy the following conditions
\be F(b)=F(-b)=0.\ee
From these conditions the energy spectrum  can be found.

In the dimensionless form equation (\ref{Shroedinger_diff}) can be written as \
\be \label{dimensionless} \left(\frac{g^2(by)}{b^2}+\varepsilon\right)f''(y)=-{\alpha_0 f(y)},\ee
with $y=\frac{p}{b},\  y\in[-1,1]$, $\varepsilon=-\frac{2mE}{b^2}$, $\alpha_0=\frac{2m\alpha}{\hbar^2}$ and $f(y)=F(by)$.
Function $f(y)$ have to satisfy the following conditions with 
\be\label{cond} f(1)=f(-1)=0.\ee
Thus, the problem is reduced to the solving of equation (\ref{dimensionless}) with boundary conditions (\ref{cond}).
\section{Energy spectrum for special cases of deformation function}
In this section we present energy spectrum of considerable problem for some special examples of deformation function.
\subsection{Example 1}
Let us consider the simplest deformed commutation relation leading to minimal length
\be \label{cutoff}
[X,P]=i\hbar, P\in[-b,b].
\ee
In this case $g(p)=p $ and equation (\ref{dimensionless}) writes
\be\label{Eq_cutoff} \left(y^2+\varepsilon\right)f''(y)=-{\alpha_0 f(y)},\ee
 Equation (\ref{Eq_cutoff}) has two linearly independent solutions finite at $y=0$, even and odd with respect to $y$:
\be\label{f_even}
&&f_{even}(y)=C_1(y^2+\varepsilon^2)\  _{2}F_{1}\left(a,b;c;-\frac{y^2}{\varepsilon}\right),\\
&&f_{odd}(y)=C_2 y(y^2+\varepsilon^2) \  _{2}F_{1}\left(a+1/2,b+1/2;c+1;-\frac{y^2}{\varepsilon}\right),
\ee

with 
\be
&&a=\frac{3}{4}+\frac{\kappa}{2},\\
&&b=\frac{3}{4}-\frac{\kappa}{2},\\
&&c=\frac{1}{2},\\
&&\kappa=\sqrt{1/4-\alpha_0}.
\ee
Here$\  _{2}F_{1}$ denotes hypergeometric function.
The energy spectrum can be found from the conditions
\be
&&\label{energy_cond_even}_{2}F_{1}\left(a,b;c;-\frac{1}{\varepsilon}\right)=0,\\
&&\label{energy_cond_odd} _{2}F_{1}\left(a+1/2,b+1/2;c+1;-\frac{1}{\varepsilon}\right)=0.
\ee

 In the case of weakly attractive potential $0<\alpha_0\leq1/4$, which correspondes to the case of  real $\kappa$, equation (\ref{energy_cond_even}) has one solution $\epsilon_0$, while (\ref{energy_cond_odd}) has no solutions (\textit{see Fig.1}). 

\begin{figure}[h]
\centering
\includegraphics[width=9 cm]{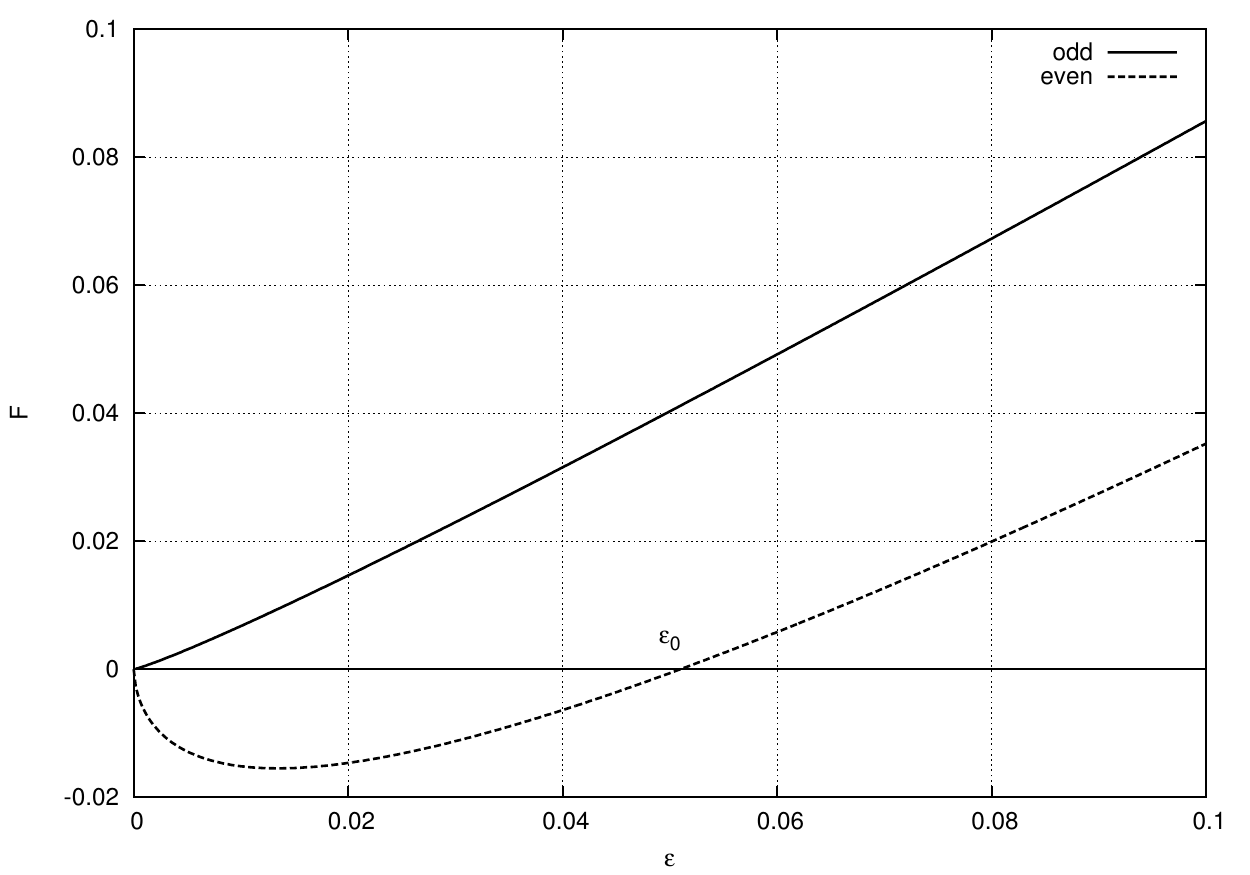}
\vspace{-10 pt}
\caption{\footnotesize{Plots of hypergeometric functions from (\ref{energy_cond_even}) and (\ref{energy_cond_odd}) corresponding to even (dotted line) and odd (solid line) eigenfunctions respectively in case of  weakly attractive potential $\alpha_0=1/4$.  }}
\label{fig1}
 \end{figure}  
 Analitical expression for $\varepsilon_0$ can be found from (\ref{energy_cond_even}) using   the following property for hypergeometric function 
 \be\label{F}
_{2}F_{1}\left(a,b;c;-\frac{1}{\varepsilon}\right)=C_1\varepsilon^{a} {}_{2}F_{1}\left(a,1-c+a;1-b+a;-{\varepsilon}\right)\nonumber\\ +C_2\varepsilon^{b}{}_{2}F_{1}\left(a,1-c+b;1-a+b;-{\varepsilon}\right),
\ee
with
\be
C_1=\frac{\Gamma(c)\Gamma(b-a)}{\Gamma(b)\Gamma(c-a)},\ \ \ C_2=\frac{\Gamma(c)\Gamma(a-b)}{\Gamma(a)\Gamma(c-b)}.
\ee
With the assumption of small $\varepsilon_0$ condition (\ref{energy_cond_even}) reads
\be\label{F_1}
_{2}F_{1}\left(a,b;c;-\frac{1}{\varepsilon_0}\right)\approx C_1\varepsilon_0^{a}+ C_2\varepsilon_0^{b}=0.
\ee
From (\ref{F_1}) we obtain the  energy 
\be\label{e_0}
\varepsilon_0=\exp\left({\frac{1}{\kappa}\ln\left(-\frac{C_1}{C_2}\right)}\right),
\ee
which is in very good coincidence with the numerical results (see Fig.2).
It is interesting that in case of infinitesimal  coupling constant $\alpha_0$ there is bound state with infinitesimal energy $\varepsilon\approx\pi^2\alpha_0^2/4$, while in case of $\alpha_0=0$ obviously there is no bound state with zero energy. This fact correspondes to the vanishing of the eigenfunction (\ref{f_even}) in the limit of $\alpha_0$ to zero.
\begin{figure}[h!]
\centering
\includegraphics[width=9 cm]{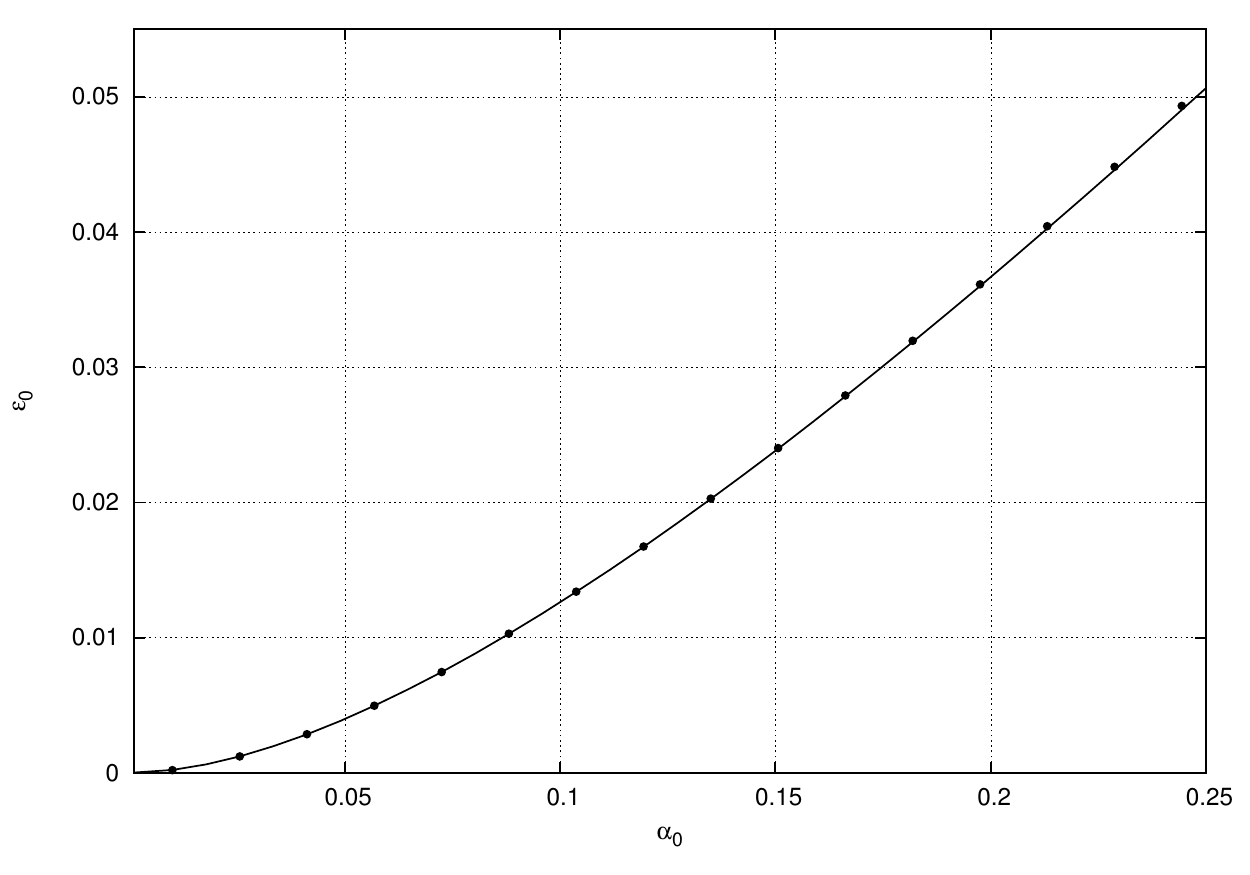}
\caption{\footnotesize{Numerical (dotted) and analytical (solid line) dependence of the energy level $\varepsilon_0$ on coupling constant $\alpha_0$} }
\label{fig1}
 \end{figure}
 
In case of strongly attractive potential $\alpha_0>1/4$, parameter $\kappa$ become complex
\be
\kappa ={i}\nu, \ \ \nu = \sqrt{\alpha_0-1/4}.
\ee
Using (\ref{F})  in the limit of small $\varepsilon$ conditions on energy spectrum (\ref{energy_cond_even}) and (\ref{energy_cond_odd})  take the form 
\be\label{two}
&&_{2}F_{1}\left(a,b;c;-\frac{1}{\varepsilon}\right)\sim_{\varepsilon\rightarrow0}\varepsilon^{3/4}\cos(-\frac{\nu}{2}\ln\varepsilon+\arg(A))=0,\\
 &&_{2}F_{1}\left(a+1/2,b+1/2;c+1;-\frac{1}{\varepsilon}\right)\sim_{\varepsilon\rightarrow0}\varepsilon^{5/4}\cos(-\frac{\nu}{2}\ln\varepsilon+\arg(B))=0.\nonumber
\ee
with
\be
&&A=\frac{\Gamma(a-b)}{\Gamma(a)\Gamma(c-b)},\\
&&B=\frac{\Gamma(a-b)}{\Gamma(a+1/2)\Gamma(c-b+1/2)}.
\ee
Both  equations (\ref{two}) has many solutions accumulating at $\varepsilon=0$. 
The energy spectrum consists of two branches corresponding to even and odd eigenfunctions 
\be
&&\varepsilon_n^{even}=\exp{\frac{2}{\nu}\left[\arg(A)-(n+\frac{1}{2})\pi\right]}, \ 
\\
&&\varepsilon_m^{odd}=\exp{\frac{2}{\nu}\left[\arg(B)-(m+\frac{1}{2})\pi\right]}, \ 
\ee
with $n, m=0,1,2,\dots$.
However this formulas  incorrectly describes the energy spectrum of the system, when the condition of smallness of the values of energy $\varepsilon\ll1$ is not fulfiled  (see Fig.3).

\begin{figure}[h!]
\centering
\includegraphics[width=6.75 cm]{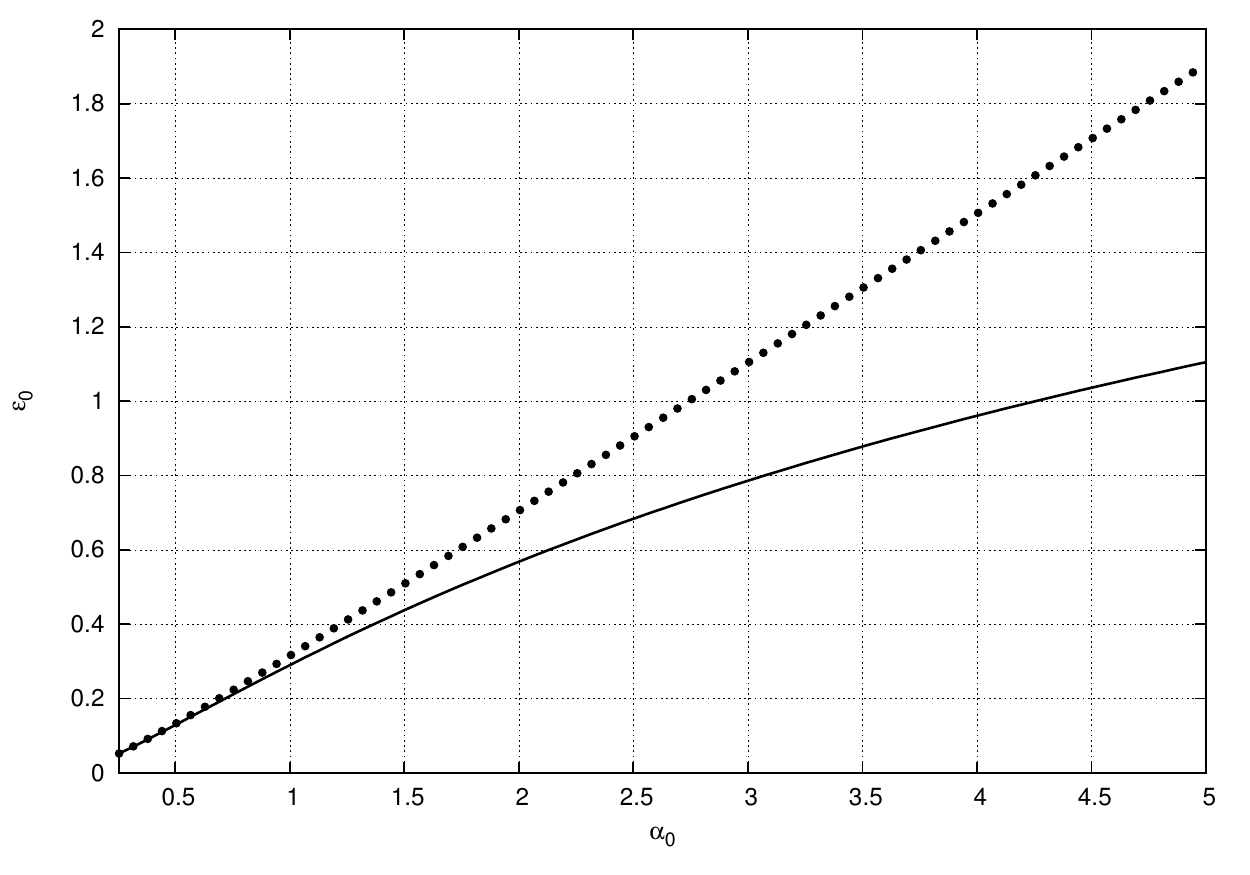}
\includegraphics[width=6.75 cm]{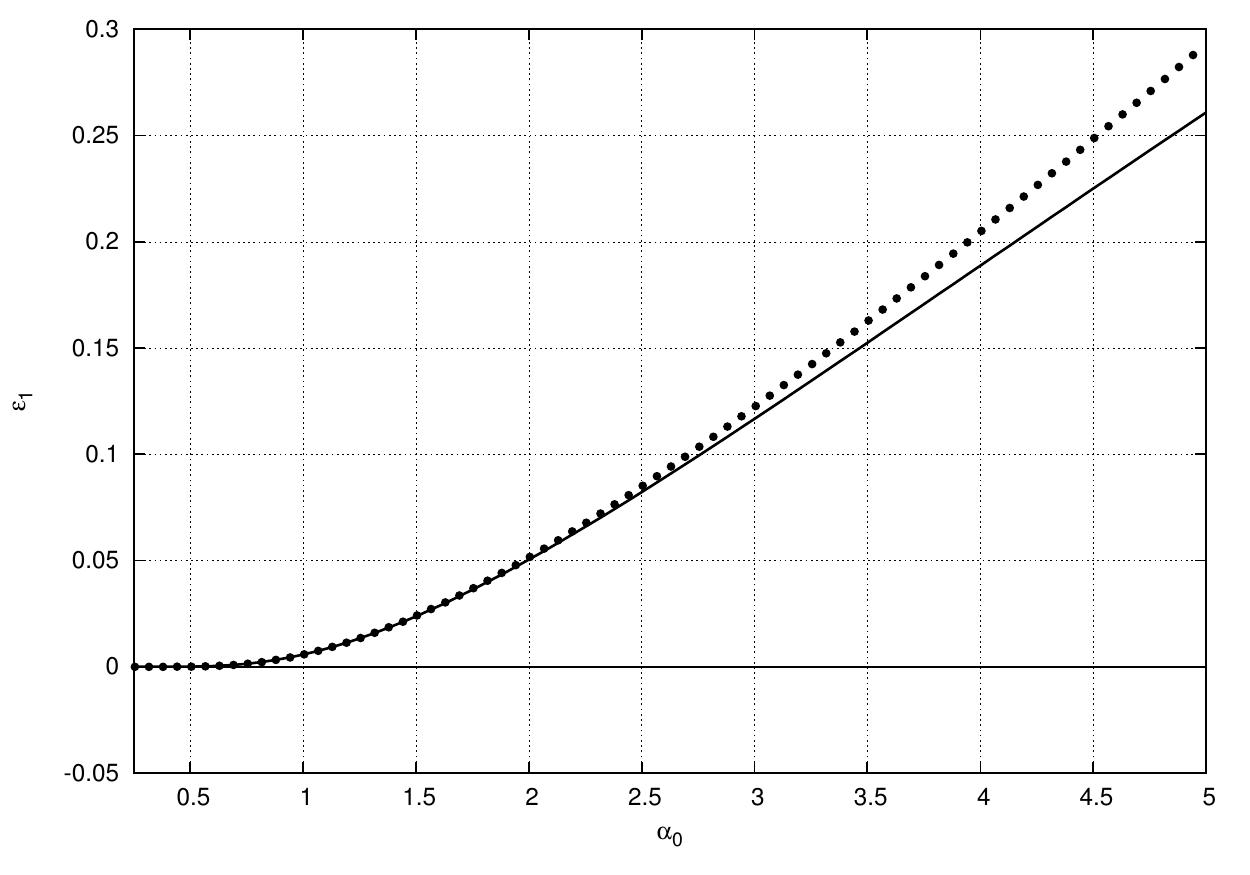}
\caption{\footnotesize{Numerical (dotted line) and analytical (solid line) dependences of the ground state energy level $\varepsilon_0=\varepsilon_0^{even}$ and the first excited state energy level  $\varepsilon_1=\varepsilon_0^{odd}$ on coupling constant $\alpha_0$.} }
\label{fig1}
 \end{figure}

Analitical behavior for large energies can be obtained from the Bohr-Sommerfeld quantization rule in deformed space with minimal length (see \cite{Fityo2008}) 
\be\label{wkb}
-\oint{XdP}=2\pi\hbar(n+\delta),
\ee  
with $n\in N$ and $\delta \in [0,1)$. For considerable system $H(X,P)=\frac{P^2}{2m}-\frac{\alpha}{X^2}$
from equation $H(X,P)=E$ we find 
\be
X=\frac{\sqrt{2m\alpha}}{\sqrt{P^2-2mE}}
\ee
Thus, we write quantization rule as 
\be\label{wkb1}
2\int_{-b}^{b}{\frac{\sqrt{2m\alpha}}{\sqrt{P^2-2mE}}dP}=2\pi\hbar(n+\delta)
\ee  
In dimensionless form (\ref{wkb1}) reads
\be\label{wkb2}
\int_{-1}^{1}{\frac{\sqrt{\alpha_0}}{\sqrt{y^2+\varepsilon}}dy}=\pi(n+\delta)
\ee 
From (\ref{wkb2}) we obtain the energy spectrum 
\be
\varepsilon_n=\frac{4 \exp(\frac{\pi(n+\delta)}{\sqrt{\alpha_0}})}{\left(\exp(\frac{\pi(n+\delta)}{\sqrt{\alpha_0}})-1\right)^2}.
\ee
In the limit of very strong attraction $\alpha_0 \rightarrow\infty$ energy spectrum linearly depends on coupling constant $\alpha_0$
\be\label{assymp}
\varepsilon_n=\frac{4\alpha_0}{\pi(n+\delta)^2}.
\ee
Comparison of the numerical results and the one's obtained from Bohr-Sommerfeld quantization rule with $\delta=0$  for ground and first excited state is presented on Fig. 4.  

\begin{figure}[h!]
\centering
\includegraphics[width=6.75 cm]{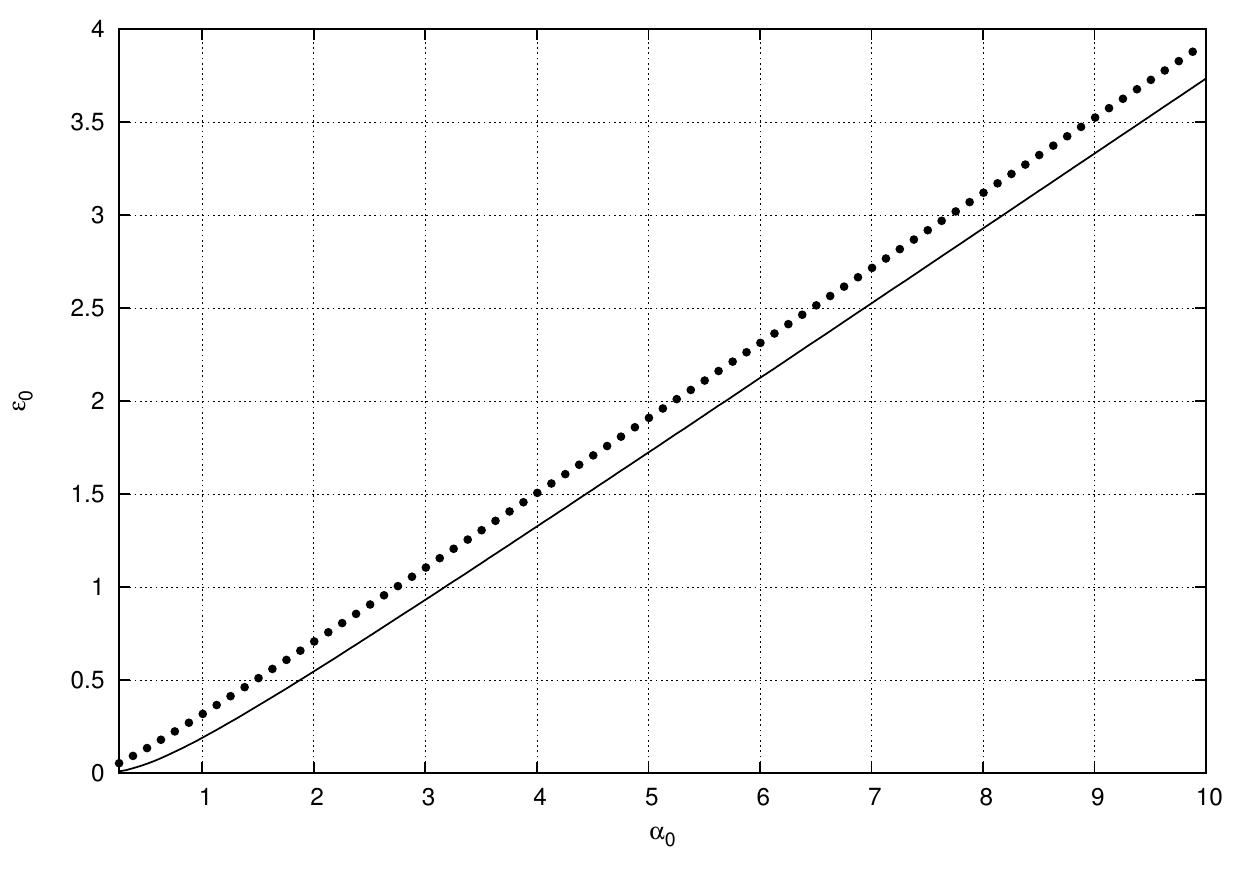}
\includegraphics[width=6.75 cm]{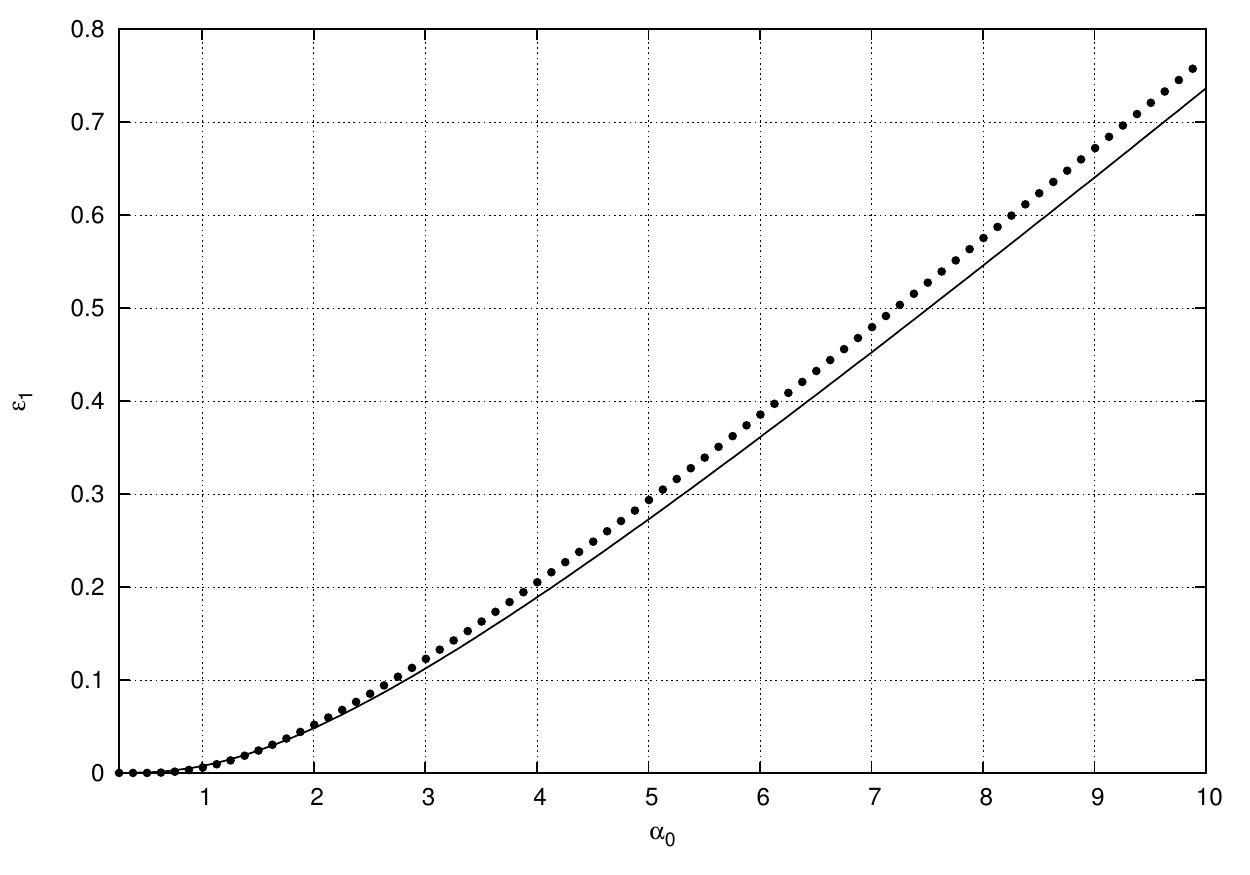}
\caption{\footnotesize{Numerical dependences(dotted line) and the analitical ones obtained from Bohr-Sommerfeld quantization rule with $\delta=0$ (solid line)  of the ground state energy level $\varepsilon_0$ and the first excited state energy level  $\varepsilon_1$ on coupling constant $\alpha_0$} }
\label{fig1}
 \end{figure}
 
 From Fig. 4 we see that asymptotic behavior at $\alpha_0 \rightarrow\infty$  of the energy spectrum obtained from WKB aproximation method  given in (\ref{assymp}) coinsides with the numerical results.
 
 Thus, one dimensional inverse square potential is regularized in deformed space with minimal length. Regularization of Coulomb potential in frame of electrodynamics in space with minimal length is presented in \cite{Tkachuk2007}.

\subsection{Example 2}
Let us consider the following deformation function
\be \label {ex2}f(P)=(1+\beta P^2)^{3/2},\ \ \ a=\infty,\\
 g(p)=\frac{p}{\sqrt{1-\beta p^2}},\ \ \ \ \ b=\frac{1}{\sqrt{\beta}}.\ee
 
 In this case   equation (\ref{dimensionless}) takes the form 
 \be \left(\frac{y^2}{1-y^2}+\varepsilon\right)f''(y)=-{\alpha_0 f(y)},\ee
Both even and odd with respect to $y$  solutions of later equation  can be presented by confluent Heun function
\be
f_{even}(y)=(y^2+\varepsilon-\varepsilon y^2)HeunC\left(0, -\frac{1}{2}, 1, \frac{\alpha_0\varepsilon}{4(1-\varepsilon)^2}, \frac{2+\alpha_0-2\varepsilon}{4-4\varepsilon}, -\frac{(1-\varepsilon) y^2}{\varepsilon}\right),\\
f_{odd}(y)=y(y^2+\varepsilon-\varepsilon y^2)HeunC\left(0, \frac{1}{2}, 1, \frac{\alpha_0\varepsilon}{4(1-\varepsilon)^2}, \frac{2+\alpha_0-2\varepsilon}{4-4\varepsilon}, -\frac{(1-\varepsilon) y^2}{\varepsilon}\right).
\ee
Energy spectrum can be found from the conditions
\be\label{2_1}
HeunC\left(0, -\frac{1}{2}, 1, \frac{\alpha_0\varepsilon}{4(1-\varepsilon)^2}, \frac{2+\alpha_0-2\varepsilon}{4-4\varepsilon}, -\frac{(1-\varepsilon)}{\varepsilon}\right)=0,\\
\label{2_2}HeunC\left(0, \frac{1}{2}, 1, \frac{\alpha_0\varepsilon}{4(1-\varepsilon)^2}, \frac{2+\alpha_0-2\varepsilon}{4-4\varepsilon}, -\frac{(1-\varepsilon)}{\varepsilon}\right)=0.
\ee

In the limit of $\varepsilon\rightarrow0$  using the following relation between hypergeom and confluent Heun function 
\be
_2F_1(\alpha, \beta; \delta, z) = \frac{1}{(1-z)^\beta}HeunC(0, \delta-1, \beta-\alpha, 0, \frac{(\beta-1+\alpha)\gamma}{2}+\frac{1}{2}-\beta\alpha, \frac{z}{z-1})
\ee
it can be shown that (\ref{2_1}) and (\ref{2_2}) coinsides with (\ref{energy_cond_even}) and (\ref{energy_cond_odd}) respectively.

Thus, the energy spectrum for this type of deformation function is close to the one obtained by a cutoff regularization. Such a conclusion can also be made on the basis of the numerical calculation (see Fig. 5)  
\begin{figure}[h!]
\centering
\includegraphics[width=6.75 cm]{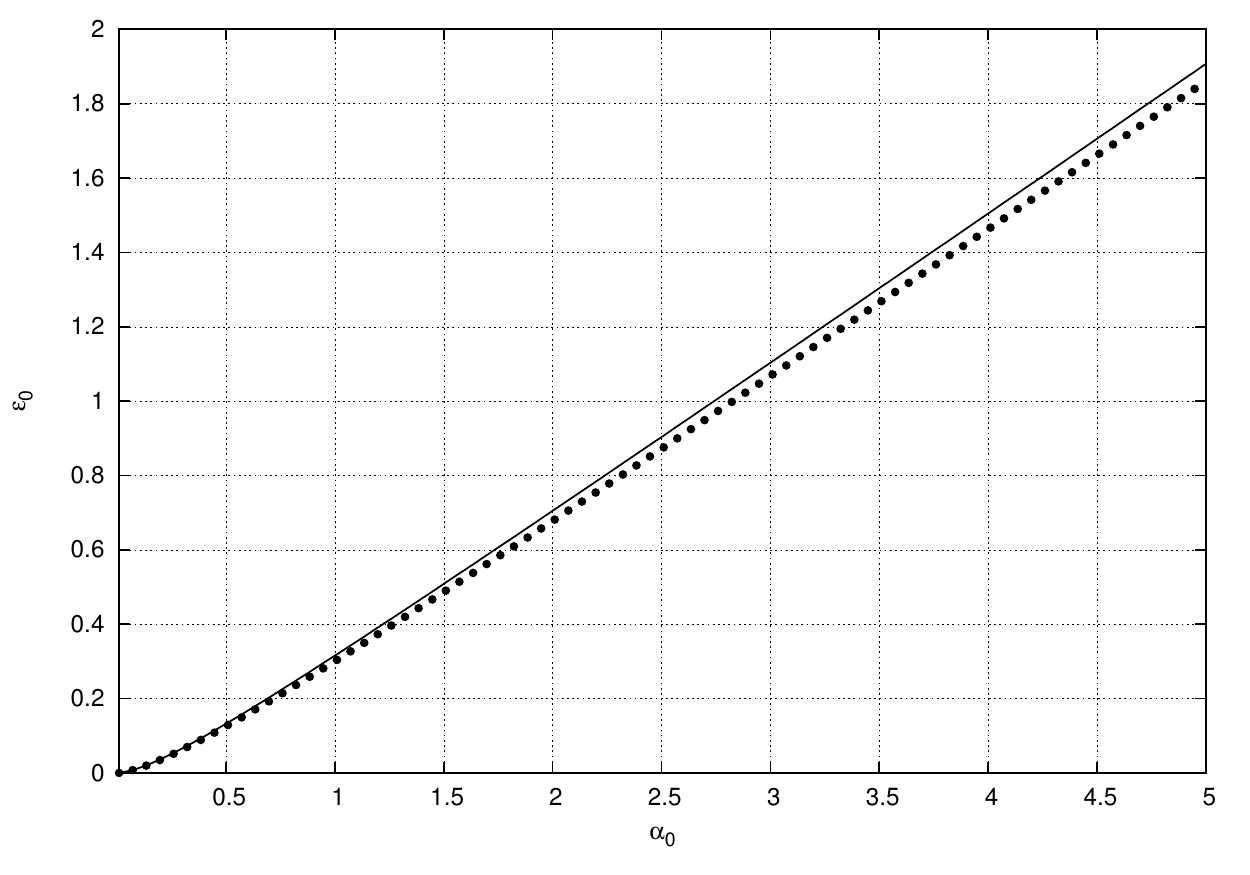}
\includegraphics[width=6.75 cm]{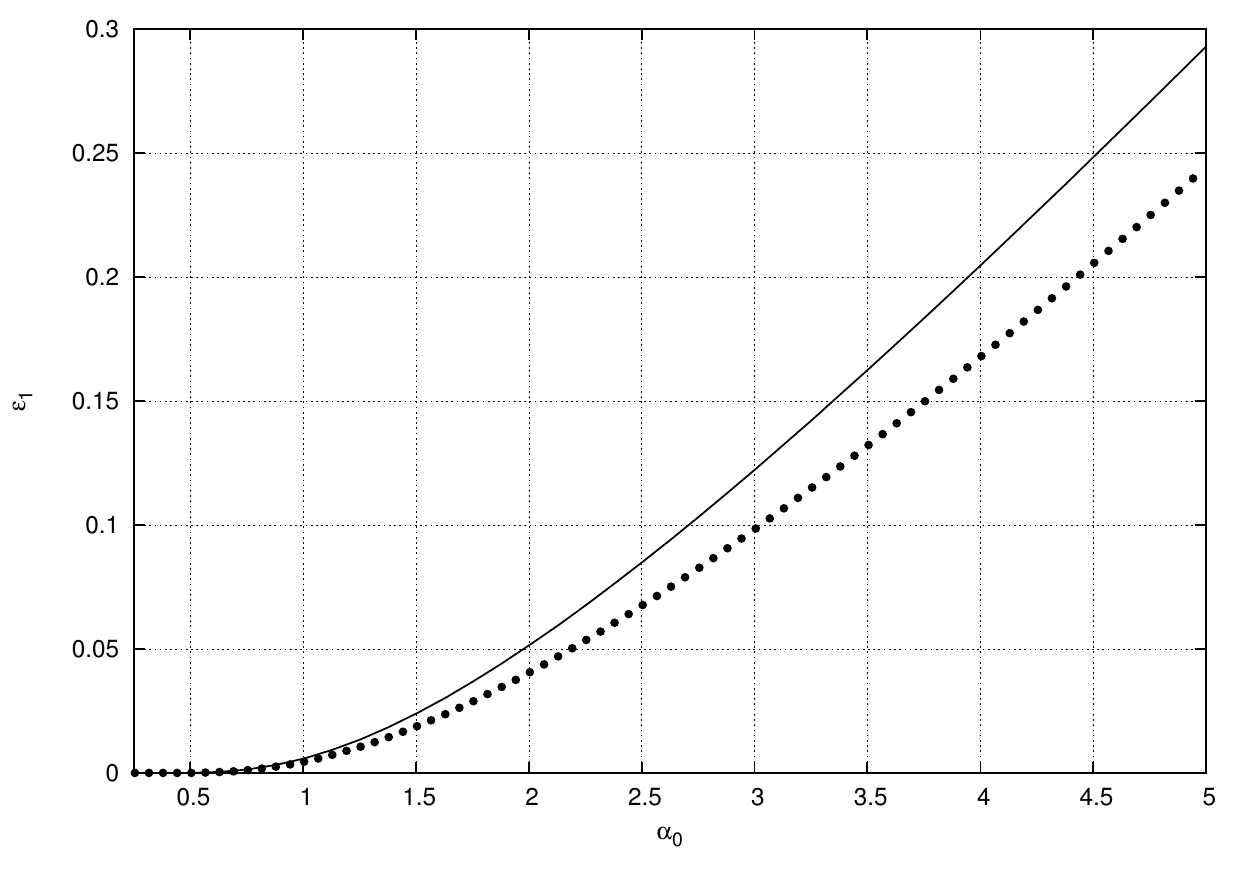}
\caption{\footnotesize{Ground state energy level and the first excited state energy level  in case of functions of deformation from the first example (solid line) and the second one(dotted line)  as function of coupling constant $\alpha_0$} }
\label{fig1}
 \end{figure}

\subsection{Example 3}
We also provide the numerical calculations of the energy spectum for the following deformation functions
 \be \label{a} a) f(P)=(1-\beta P^2)^{1/2},\  a=\frac{1}{\sqrt{\beta}};\ \ g(p)=\frac{1}{\sqrt{\beta}}\sin(\sqrt{\beta}p),\  b=\frac{\pi}{2\sqrt{\beta}};\\ \label{b} b) f(P)=(1+\beta P^2),\ \ \ \  a=\infty;\ \ \ \ 
 g(p)=\frac{1}{\sqrt{\beta}}\tan(\sqrt{\beta}p),\  b=\frac{\pi}{2\sqrt{\beta}};\ee 
 and provide the comparison with the corresponding results for the first example of function of deformation.
 \begin{figure}[h!]
\centering
\includegraphics[width=6.75 cm]{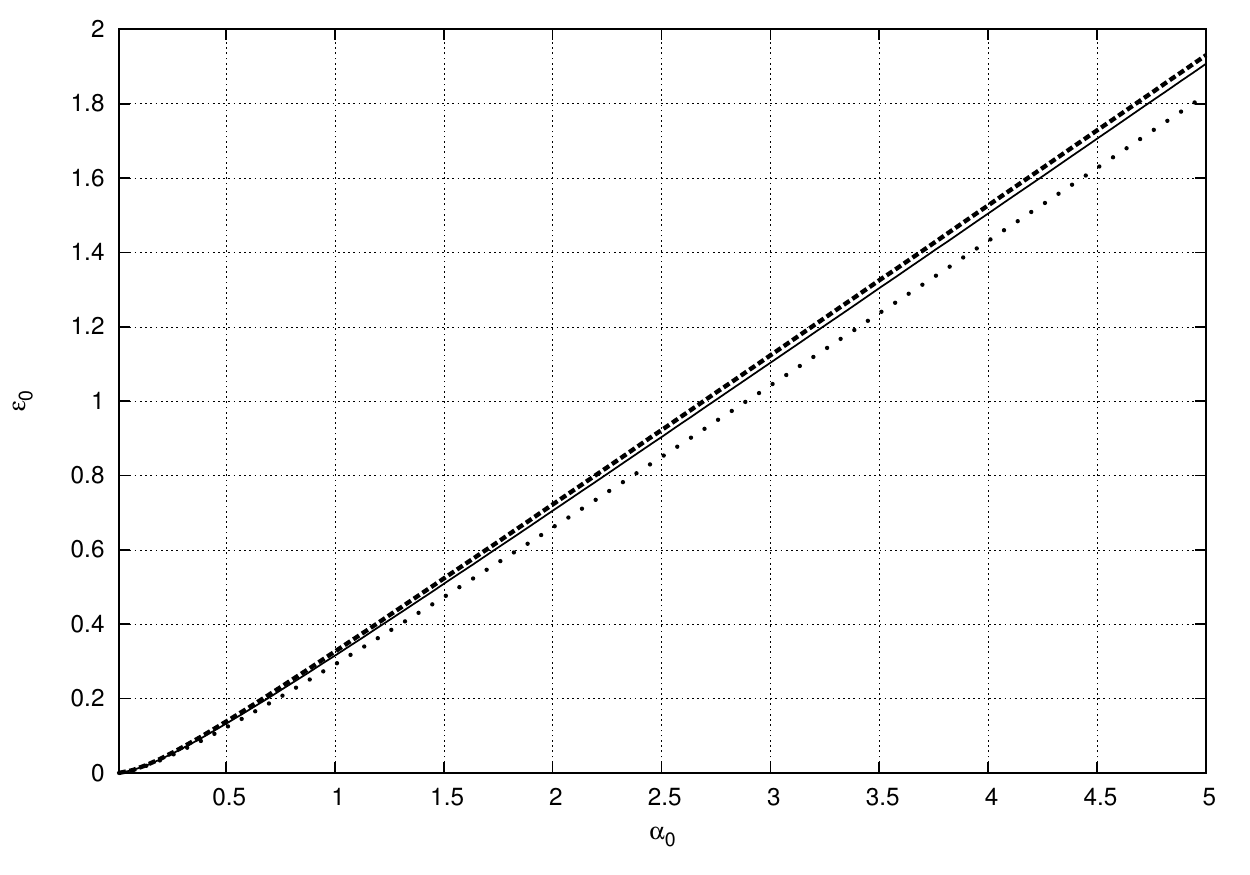}
\includegraphics[width=6.75 cm]{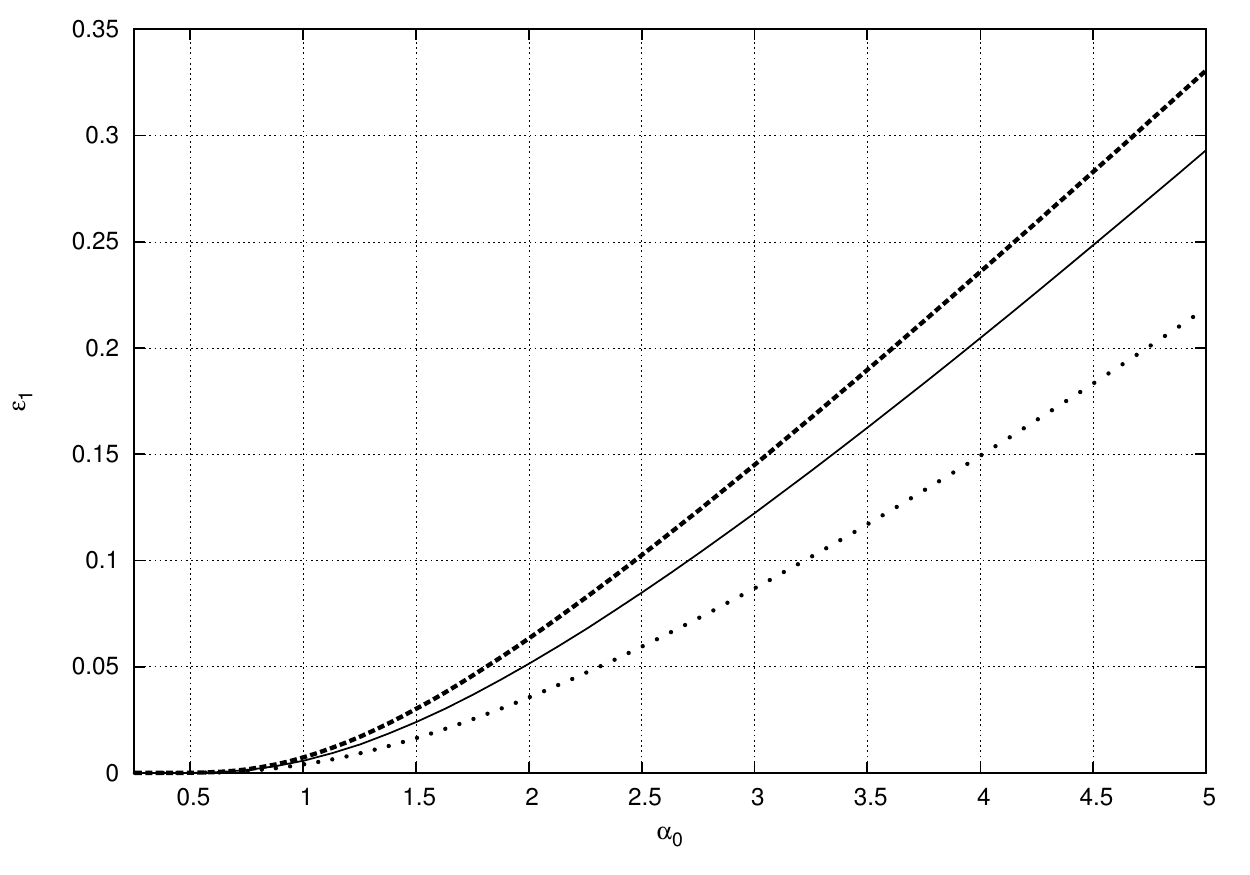}
\caption{\footnotesize{Ground state energy level and the first excited state energy level  in case of functions of deformation from the  example 2 (solid line),  3a (dashed line) and 3b  (dotted line)  as function of coupling constant $\alpha_0$}}
\label{fig1}
 \end{figure}
As we can see from Fig. 6 the  values of the energy for the ground state and the first exited state slightly depends on the choice of deformation function in case of coupling constant $\alpha_0<1$. This conclusion is valid also for the case for the function of deformation considered in Example 2. Ground state energy is almost the same for considered types of deformations (including Example 2) in wide range of 
 coupling constant. The maximal difference of the ground state energies is $\frac{\varepsilon_{max}(\alpha_0)-\varepsilon_{min}(\alpha_0)}{\varepsilon_{min}(\alpha_0)}<6\%$ for any choice of $\alpha_0$.
 
\begin{figure}[h!]
\centering
\includegraphics[width=9 cm]{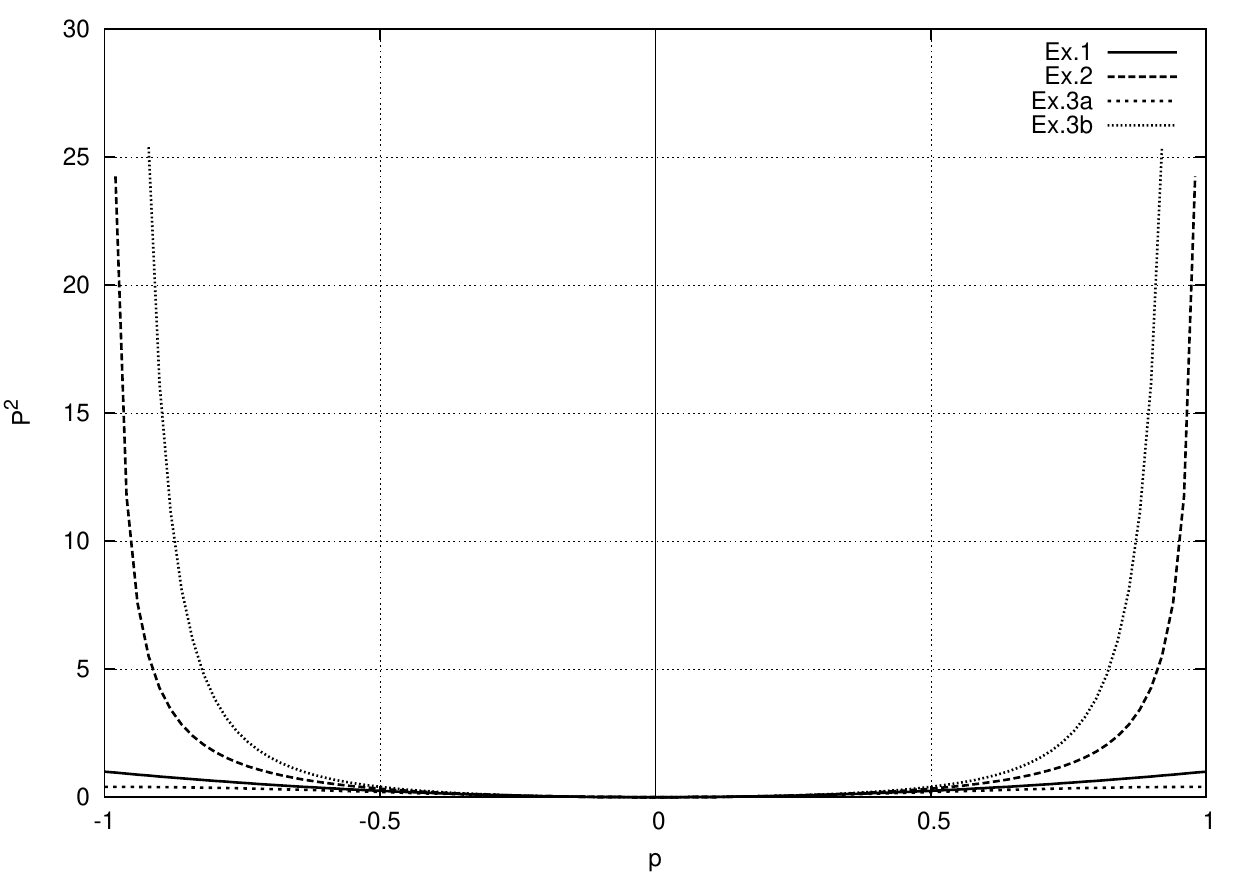}
\caption{\footnotesize{Plots of kinetic energies  $g^2(p)$ for the considered examples of deformed space}}
\label{fig7}
 \end{figure}

 Let us consider  two Hamiltonians of the particle in the inverse square  potential for two different derormed space.
 \be
 H_a=\frac{g^2_a(p)}{2m}-\frac{\alpha}{X^2}, \\
 H_b=\frac{g^2_b(p)}{2m}-\frac{\alpha}{X^2},
 \ee
 with $p\in[-b,b]$ for both the Hamiltonians.
 Howewer the definition of  the inverse square  potential does not depends on the choice of deformation function $g(p)$ we conclude that $\langle\phi|H_b|\phi\rangle>\langle\phi|H_a|\phi\rangle$ if $g^2_b(p)>g^2_a(p)$ for all the values of $p$. If we consider ground state eigenfunction $\phi_b$ of the Hamiltonian $H_b$ we can write for the energy of ground state of  the Hamiltonian $H_b$ the following relation $E_b=\langle\phi_b|H_b|\phi_b\rangle>\langle\phi_b|H_a|\phi_b\rangle$. But we know that $\langle\phi_b|H_a|\phi_b\rangle>\langle\phi_a|H_a|\phi_a\rangle=E_a$, with $\phi_a$ being the  
ground state eigenfunction of the Hamiltonian $H_a$.   Finally we conclude that  if $g^2_b(p)>g^2_a(p)$ for all the values of $p$ the following condition is hold $E_b>E_a$. 
This result can also be applicable for excited states.
On Fig. (\ref{fig7}) we present plots of kinetic energies  $g^2(p)$ for the consedered examples of deformed space. We see that among considered deformed spaces the one given in Ex. 3b  and Ex. 3a have the biggest and the smallest ground state energy correspondingly. This results is also confirmed by numerical calculations.

\section{Conclusion}

In this paper we have studied the general case of deformed Heisenberg  algebra leading to the minimal length.  We  have proposed the definition of square inverse position operator  (\ref{X^-2}). This definition has been obtained from the functional analysis of the square position operator in deformed space with  minimal length. Using definition of the square inverse position operator  we have studied quantum particle in  the field of inverse square potential in the general case of deformed Heisenberg  algebra leading to the minimal length.  The problem has been reduced to solving of the equation (\ref{dimensionless}) with boundary condition (\ref{cond}).

Particular examples of the deformation function have been studied. 
In the simplest case of deformation function (\ref{cutoff}), which corresponds to cutoff procedure in momentum space, we have obtained the eigenfunctions in terms of hypergeometric function and transcendental equation for energy spectrum. 
 In the case of weakly attractive potential $0<\alpha_0\leq1/4$  energy spectrum consists of one energy level  (\ref{energy_cond_even}) corresponding to even eigenfunction with respect to momentum. We have derived the aproximate analytical formula (\ref{e_0}) for this energy level  which is in very good coincidence with the numerical calculations. It is interesting that in case of infinitesimal  coupling constant $\alpha_0$ there is bound state with infinitesimal energy, while in case of $\alpha_0=0$  there is no bound state with zero energy. 
In case of strongly attractive potential $\alpha_0>1/4$ the energy spectrum consists of two branches of energies accumulating at $\varepsilon=0$   corresponding to even and odd eigenfunctions.  We have obtained  aproximate analytical formulas for this two branches in the assumption of small energies. From the  numerical solutions for energy spectrum we have concluded that this approximation is valid in case of  $\alpha_0<1$. From numerical calculation 
it have been shown the linear dependence of the energies on coupling constant $\alpha_0$ in the limit of big $\alpha_0$. The same dependence have been obtained analytically  from 
 the  Bohr-Sommerfeld quantization rule for deformed space with minimal length. 
 
 In the second example of deformation function (\ref{ex2}) we have obtained the eigenfunctions in terms of confluent Heun function and transcendental equation for energy spectrum. 
 In the approximation of small energies we have shown that energy spectrum for deformation function (\ref{ex2})  coinsides with the one obtained in the  case of deformation function (\ref{cutoff}). This conclusion have been confirmed by numerical calculation.
 
 Finally numerical solution of the equation (\ref{dimensionless}) with boundary condition (\ref{cond}) have been obtained for two more examples of deformation functions (\ref{a}) and (\ref{b}). The  values of the energy for the ground state and the first exited state slightly depends on the choice of deformation function in case of coupling constant $\alpha_0<1$. This conclusion is valid for all considered examples  of the functions of deformation. Ground state energy is almost the same for considered types of deformations in wide range of 
 coupling constant. The maximal difference of the energies does not exceed $6\%$.
 
\section{Acknowlegments}
This work was partly
supported by the grant of
the President of Ukraine for support of scientific researches of young scientists (F-75).

\end{document}